\documentclass[aps,prl,twocolumn,amsfonts,showpacs,superscriptaddress]{revtex4} 
\pdfoutput=1
\usepackage{bm}
\usepackage{epsfig,amsopn}
\usepackage{graphicx}
\usepackage{amsmath,amssymb}
\usepackage{natbib}
\usepackage{color}
\renewcommand{\section}[1]{{\par\it #1.---}}
\def\etal{{\emph{et al~}}}
\def\ql{ {Q_{ l}} }
\def\qr{ {Q_{ r}} }

\newcommand \bea{\begin{eqnarray}}
\newcommand \eea{\end{eqnarray}}
\newcommand \f{\frac}
\newcommand \nn{\nonumber}
\newcommand \ra{\rangle}
 \newcommand \la{\langle}
\newcommand \p{\partial}
\begin{document}
\title{ Exact solution of a Levy walk model for  anomalous heat transport}
\author{Abhishek Dhar}
\affiliation{International  Centre for Theoretical Sciences, TIFR, Bangalore 560012, India}

\author{Keiji Saito}
\affiliation{Department of Physics, Keio University, Yokohama 223-8522, Japan} 

\author{Bernard Derrida}
\affiliation{
 Laboratoire de Physique Statistique, Ecole Normale Sup\'erieure,
UPMC, Universit\'e Paris Diderot, CNRS,
24 rue Lhomond, 75231 Paris cedex 05, France}
\date{\today} 

\begin{abstract}
{
The Levy walk model is studied in the context of 
the anomalous heat conduction of one dimensional systems.
In this model the heat 
carriers execute Levy-walks instead of normal diffusion as expected in systems 
where Fourier's law holds. Here we 
calculate  exactly  the average 
heat current, the large deviation function of its fluctuations and  the 
temperature profile of the Levy-walk model  maintained in a steady state by contact with two heat baths (the open geometry). 
We find that  the  current is  non-locally connected to the temperature gradient.
 As observed  in recent simulations of mechanical models,   all  the  cumulants   of the  current fluctuations have  the same system-size dependence in the open geometry. 
 For the ring geometry, we   argue that a size dependent cut-off time is necessary for the Levy walk model to behave as mechanical models.
This modification does not affect the results on transport in the open geometry  for large enough system sizes.}
\end{abstract}

\maketitle
\newpage

\section{Introduction}
Fourier's law of heat conduction (in one dimension) states that 
\bea
J(x,t)= -\kappa \f{\p T(x,t)}{\p x}, \label{foureq}
\eea
where $T(x,t),~J(x,t)$  are  the local temperature and    heat current density fields and $\kappa$ is the thermal conductivity. 
Based on results obtained from a  large number of numerical simulations 
and various analytical approaches  it is now believed that Fourier's law is not valid in one and two dimensional mechanical systems  (in particular  models where momentum is conserved) and heat conduction is anomalous \cite{BLR00,LLP03,dhar08}.  
Anomalous behavior in heat conduction is not only a theoretical issue but recently of experimental relevance in  
several low-dimensional materials \cite{chang08,nika09}.
Indicators of the anomaly include --- (i) in steady states the dependence of 
the  heat  current $J$ on  system size $L$ shows the scaling behaviour $J \sim L^{\alpha-1}$ with 
$\alpha > 0$, (ii) the temperature profiles across systems in  
nonequilibrium steady states are found to be nonlinear, even for very small applied temperature differences 
and, (iii) the spreading of heat pulses in anharmonic chains is super-diffusive. 

The microscopic basis of Fourier's law lies in the fact that the carriers of 
heat in a system execute random walks.  The simplest ``derivation'' of 
Fourier's  law, from kinetic theory, is based on this picture and leads to
an expression of the conductivity of a material in terms of the mean free 
path $\ell$, mean velocity $v$ and specific heat capacity $c$ of the heat 
carriers: $\kappa= c v \ell$ (in one dimensions). The breakdown of Fourier's 
law thus also implies a breakdown of the random walk picture and of the diffusion equation 
\bea
\frac{\p T(x,t)}{\p t}=\frac{\kappa}{c} \frac{\p^2 T(x,t)}{\p x^2} ~,\label{diffeq}
\eea
which describes time-dependent heat transfer (assuming $\kappa$ has no 
temperature-dependence). A number of recent 
studies indicate that  a  good description of anomalous heat conduction in
one dimensional systems is obtained by modeling the motion of the heat carriers
as  Levy random walks instead of simple random walks \cite{denisov03,metzler04,cipriani05,lepri11}. 
Numerical studies show that the spreading of heat pulses 
and the form of steady state temperature profiles can be correctly modeled 
by means of the Levy walk. In another study of a model 
with stochastic dynamics, it has been shown, starting from a   
Boltzmann-equation approach, that the temperature satisfies a fractional 
diffusion equation corresponding to a Levy stable process \cite{jara09}. 

While some analytical understanding has been achieved \cite{denisov03} it is 
desirable to further develop the Levy walk theory for anomalous heat conductivity so that (i) one can use it in the way as one uses Eqs.~(\ref{foureq},\ref{diffeq}) for normal diffusion and (ii) one has a clearer idea of the range of applicability of the model. 
In this Letter we present 
several exact results for steady state heat transport in the  Levy walk model in one dimension.
We obtain exact analytic expressions for properties  such as the density profiles or the average current, which agree with what was already known numerically.
We also obtain new results for properties such as the cumulants of the current fluctuations which had not been considered before. 
For setting up the steady state we follow the idea in \cite{lepri11} of connecting  two infinite 
reservoirs at different temperatures to the system and consider a version 
where space and time are taken to be continuous.

Our exact results provide several interesting physical perspective on 
anomalous heat transport.
The analytic solution of the particle-density profile exhibits the 
nonlinear (and singular) form typical of temperature profiles in $1D$ systems \cite{LLP03,dhar08}.
 The steady state current has the power law dependence on the system-size,
characteristic of anomalous diffusion.
Also in
 contrast to  (\ref{foureq}) for normal heat transport, the current 
is non-locally connected to  the  temperature gradient.
In addition, we derive the exact cumulant generating function of current for the open geometry. Our results  show 
that all  cumulants  of the  integrated 
current have the same system-size dependence as the average current. This is consistent with recent 
numerical results of heat conduction in hard particle systems \cite{brunet10}
 and therefore 
  strongly suggests that the Levy-walk model gives a good description of anomalous heat conduction 
not only for the average current but also  the size dependence of the  current fluctuations. 
For the ring geometry, also, the size dependence of the cumulants obtained in  simulations \cite{brunet10}  of mechanical systems can be recovered by introducing a size dependent cut-off in the distribution of times of the Levy-walk model.

\section{Levy diffusion on the infinite line} 
In the simplest description we think of  energy  in the system as being  transported by particles performing Levy walks, 
 each particle carrying  a single quantum of energy. Therefore the local energy 
density and energy current at  any point  are  directly  proportional
to the particle density and current respectively.  In this model      
the temperature is proportional to the energy density and hence to the   
density of particles.
The precise definition of the Levy walk model that we consider here is as follows. For a single particle each step of the walk consists in choosing a time of 
flight  $\tau$ from  a given  distribution  $\phi(\tau)$ and then moving  it at  speed $v$ over a distance  $x=v \tau$ in either direction, with equal probability. Let us define 
$P(x,t) dx$ as  the probability 
that the particle is in the interval  $(x,x+dx)$ at time $t$. Thus $P(x,t)$ 
includes events where the particle is crossing the interval $(x,x+dx)$. 
If a particle starts at the  origin at time $t=0$, the probability  $P(x,t)$ satisfies
\bea
&& P(x,t)= {1 \over 2}\psi(t) \delta(|x|- v t )  \label{P(x,t)}
\\ 
&& + {1 \over 2} \int_0^t   d \tau \phi(\tau)  [P(x- v \tau,t-\tau) + P(x+ v \tau,t-\tau) ]~,
\nn
\eea
where  $\psi(\tau)=\int_\tau^\infty d\tau'~\phi(\tau')$ is the probability of choosing a time of flight $\geq \tau$.
The Fourier-Laplace transform of $P(x,t)$  can be calculated (see supplementary material \cite{suppl}) from (\ref{P(x,t)}) in terms of the Laplace transform of the distribution  of  flight times $\phi(\tau)$ and this gives the time dependence of  all the cumulants $\la x^n \ra_c$  of the position $x$ at time $t$.

If  the first moments  of the flight times  $ \langle \tau^n \rangle $ were finite, 
the motion would be  diffusive. One would  then get  from (\ref{P(x,t)}) for the first cumulants of $x$ at large $t$
\bea
 { \la x^2 \ra_{{c}}  \over  v^2 ~t}  \simeq {\la \tau^2 \ra \over \la \tau \ra}  
    ~ ; ~~  { \la x^4 \ra_c \over v^4 ~ t}  \simeq 
 3 { \la \tau^2 \ra^3  \over \la \tau \ra^3}- 6 { 
 \la \tau^2 \ra 
 \la \tau^3 \ra   \over \la \tau \ra^2} + 
{\la \tau^4
\ra \over \la \tau \ra}  ~ .
\label{dif-cum}
\ \
\eea
Here we   consider Levy walkers with a time-of-flight distribution decaying like    a power law   at large time
\bea
 \phi({ \tau}) \simeq A ~\tau^{-\beta-1} ~,~~1<\beta<2~,
\label{phiform}  
\eea
(for example  $\phi({ \tau}) = \beta /t_o / (1+{ \tau }/t_o)^{\beta+1}$ which is used for the data shown in figure 1).
 For  this  range of $\beta$  the mean flight time 
 $\la \tau \ra =\int_0^\infty d\tau ~\tau ~ \phi(\tau) $ is finite but   $\langle \tau^2 \rangle = \infty$.
Using again the Fourier-Laplace transform of   (\ref{P(x,t)})   one   gets
 for large $t$ 
\bea
\la x^2 \ra_{{c}}  &\simeq &   {2~ A ~ v^2\over (3-\beta)(2-\beta) \beta   ~ \la \tau \ra} \,  t^\gamma,~~~~\gamma={3-\beta}~, \label{xsq}   \\ 
\la x^4 \ra_c  
&  \simeq &   
{  
{4~ A ~ v^4 \over (5-\beta)(4-\beta) \beta   ~ \la \tau \ra} \,  t^{\gamma+2} ~.
} 
 \label{xqd}
\eea 
 We see that for $1<\beta<2$  the motion is  superdiffusive  \cite{zumofen93,metzler99}. 
 The solution of (\ref{P(x,t)})  
 corresponds to a pulse whose central region is a Levy-stable distribution with a scaling $ x \sim t^{1/\beta}$, has ballistic peaks  of 
magnitude $t^{1-\beta}$ at $x=\pm vt$  and vanishes outside this \cite{cipriani05}. 
These expressions (\ref{dif-cum}), (\ref{xsq}) and (\ref{xqd}) are also important for discussing current fluctuations in the ring geometry
addressed later.

\section{Levy diffusion in a finite system connected to infinite reservoirs} 
 Let us now turn to the more interesting case of the open geometry where  the system is    a  finite segment between $(0,L)$ 
connected on  its  two sides to reservoirs.
In addition to the probability density $P(x,t)$ let us also define the quantity $Q(x,t) dx dt$ as the probability that a particle has precisely landed 
in the interval $(x,x+dx)$ during the time interval $(t,t+dt)$. 
Note that at any given time, a particle could either have landed at a point $x$ or could be passing over the point. 
We need to  set up the correct boundary conditions required to construct 
a nonequilibrium current carrying  steady state. 
To do so we identify the region $  x \le 0$   with the left reservoir 
and the region $x \geq L $ with the right reservoir. We set $Q(x,t)=\ql$ for 
points on the left reservoir and $Q(x,t)=\qr$ for those on the right. 
 In the steady state we have $Q(x,t)=Q(x)$ and $P(x,t)=P(x)$, and they satisfy 
 (see supplementary material \cite{suppl})
\bea
&& Q(x)-\int_0^L  dy~\f{1}{2v} \phi(|x-y|/v) ~Q(y) \nn \\
&&~~~~~~~~=  \f{\ql}{2} \psi(x/v)
+ \f{\qr}{2} \psi[(L-x)/v]~,  \label{QeqSS} \\ 
&& P(x)= \int_0^L  dy \f{1}{2v} \psi(|x-y|/v)~ Q(y)  \nn \\
&&~~~~~~~~+  \f{\ql}{2} \chi(x/v) + \f{\qr}{2} \chi[(L-x)/v] ~. \label{PeqSS}
\eea
 where,   as in (\ref{P(x,t)}),  $\psi(t)=\int_t^\infty d\tau ~\phi(\tau )$ and $\chi(t ) =  \int_{t}^\infty d \tau \psi(\tau)$.   In the above expressions $Q(x)$ gets contributions from walkers starting from
all possible points $y$ and landing precisely at $x$. On the other hand $P(x)$ gets contributions  from walkers starting at $y$ and being either at or  passing $x$ at time.
The  problem  is closely related to the escape probability \cite{buldyrev01} of a Levy walker on the interval $(0,L)$:
if $H(x)$ is the probability that a Levy walker starting at position $x$ will first hit the left reservoir (i.e. the region $x < 0$) before it hits the right reservoir (i.e. $x >L$) it is easy to see that $ H(x)$ satisfies 
\bea
 H(x)=\int_0^L  dy~\f{1}{2v} \phi(|x-y|/v) ~H(y) 
+ \f{1}{2} \psi(x/v)
~,  \label{H(x)} 
\eea
and from  (\ref{QeqSS}), one can see that \mbox{$Q(x) = (\ql-\qr)  H(x) + \qr$}.
 
\begin{figure}
\includegraphics[width=7.5cm]{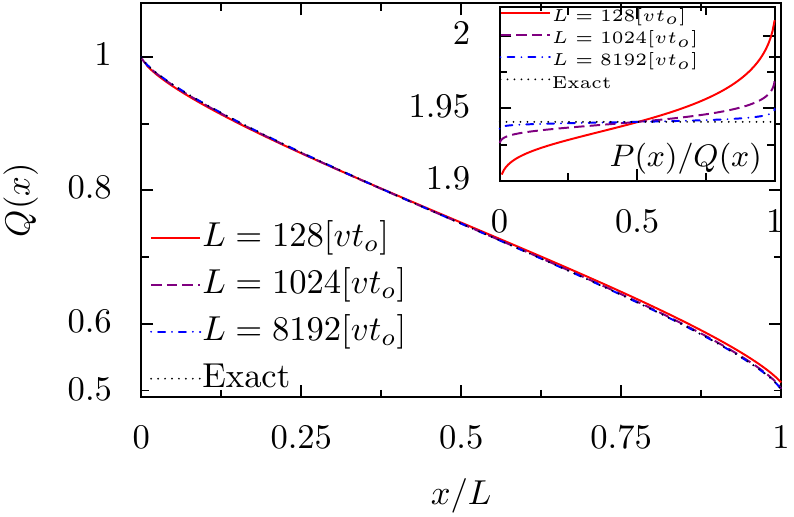}
\caption{Plot of $Q(x)$  for different system sizes, for the Levy walk with $\beta = 1.5$, $\ql=1.0, \qr= 0.5$. The data are obtained by solving Eqs.~(\ref{QeqSS},\ref{PeqSS}) with discretized space.  The distribution of flight times is $\phi (\tau)=\beta/t_o/(1+\tau/t_o)^{\beta +1}$.
The inset shows $P(x)/Q(x)$, which converges to $\chi (0)$ in the limit of $L\to\infty$.
}
\label{fig1}
\end{figure}

If one considers a Levy flight with distribution   $\rho(z)=[\phi(z/v)+\phi(-z/v)]/(2 v)$  of steps $z$, the probability $H(x)$   that  starting at $x$, the  flight hits first the left 
bath satisfies exactly Eq.~(\ref{H(x)}).  Hence
 by following the same mathematical steps as in     \cite{buldyrev01} to study equations such as (\ref{QeqSS}) or (\ref{H(x)}), one  can show  that, in the large $L$ limit,  the solution  $Q(x)$ of (\ref{QeqSS})   (and $H(x)$ of (\ref{H(x)}))  satisfies
\bea
\int_0^L dy~\psi (|x-y|/v)~ {\rm Sgn}(x-y) Q'(y)= 0~.  \label{Qpeq}
\eea
 with $Q(0)=\ql$ and $Q(L)=\qr$ (and $H(0)=1$ and $H(1)=0$ for  (\ref{H(x)})) with a  solution of (\ref{Qpeq}), for  a $\phi(\tau)$    decaying as in (\ref{phiform}), which satisfies 
\bea 
Q'(x)=-B [x (L-x)]^{\beta/2-1}~. \label{Qpsol}
\eea
Integrating this and imposing the boundary conditions $Q(0)=\ql$ and 
$Q(L)=\qr$,  one obtains the constant  $B= (\ql-\qr) ~
\Gamma(\beta)/  
\Gamma(\beta/2)^2~
L^{1- \beta}$ and  therefore $Q(x)$.  One can also notice that in (\ref{PeqSS}) the r.h.s. is dominated, for large $L$,   by the range $| y-x|\ll L$ and therefore
\bea
P(x)=\chi(0)Q(x) =  \langle \tau \rangle Q(x) ~. \label{pq}
\eea 
In Fig.~(\ref{fig1}), we  compare  numerical results obtained by solving Eqs.~(\ref{QeqSS},\ref{PeqSS}) 
 with the exact results of  Eqs.~(\ref{Qpsol},\ref{pq}). The  profiles are nonlinear and  
 look   similar to those observed for temperature profiles in 1D heat conduction \cite{LLP03,dhar08}. 

We next discuss the  current.  The steady state current $J(x)$ at position $x$ is given by (see supplementary material \cite{suppl})
\bea
 J(x) = \f{1}{2}\int_{-\infty}^{\infty} dy  ~ Q(x-y) 
 ~{\rm Sgn} (y) ~ \psi(|y|/v) ~, \label{currexp}
\eea
which can be interpreted as the difference  between  the flow from left to right and from right to left.
%%%
 The contribution to the integral coming from $y >0$ corresponds to 
particles crossing the point $x$ from left to right -- and is obtained by taking the density of particles at $x-y$ and multiplying by the probability $\psi(y/v)$  that these have a flight time longer than $y/v$. Similarly 
the other part of the integral (from $y<0$) corresponds to a right-to-left 
current. 
After a 
partial integration  and using the fact that
$Q(0) = \ql$ and $Q(L) = \qr$, one gets 
\bea
J(x)&=& -\f{v}{2} \int_{0}^{L} dy~ \chi (|x-y| / v)~ Q'(y)~. \label{jss}
\eea
We note that $dJ/dx=0$ gives Eq.~(\ref{Qpeq}) and so the current is independent of $x$, as  expected.   
Evaluating the current at $x=0$ and  using  
Eq.~(\ref{Qpsol}), we get for large $L$ 
\bea
J \simeq   ( \ql- \qr)  \  
{ A~ v^{\beta} 
~\Gamma(\beta )~ \Gamma(1- {\beta \over 2}) \over 2 ~\beta (\beta-1)~\Gamma({\beta \over 2}) 
}   \  L^{\alpha -1}  ,~~~\alpha= 2-\beta .\nonumber \\
 \label{Jsol}
\eea
From Eq.~(\ref{xsq}) we then get the relation $\alpha=\gamma-1$, 
 between the conductivity exponent 
of anomalous transport and the exponent for 
Levy-walk diffusion. This relation for Levy diffusion was noted in 
 \cite{denisov03}, numerically observed in $1D$ heat conduction models \cite{cipriani05,zhao} and a derivation based on linear response theory has recently been proposed  \cite{liuli}.

In the  large $L$ limit by using Eq.~(\ref{pq}) in Eq.~(\ref{jss}) 
we obtain 
\bea
J&=&-\f{v}{2 \la  \tau \ra }\int_0^L d{ y}~ \chi(|x - y| /v) P'(y )~. 
\eea
This is the analogue of Fourier's Law Eq.~(\ref{foureq}) in the case of normal heat conduction 
and  can be interpreted as  current being  non-locally  connected to the temperature gradient.

\section{Current fluctuations in the open system}
In the rest of this letter, we discuss current fluctuations.

\begin{figure}
\includegraphics[width=7.5cm]{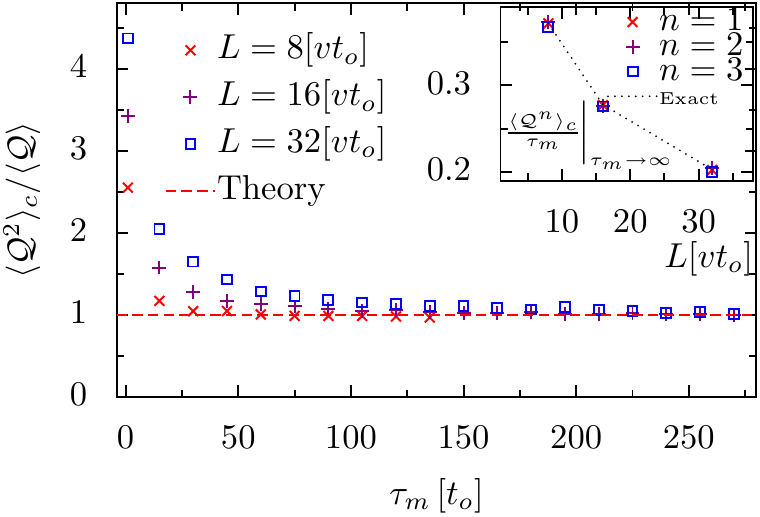}
\caption{Monte-Carlo results for $\langle {\cal Q}^2 \rangle_c / 
\langle {\cal Q} \rangle$ as a function of measurement time $\tau_m$
for several system sizes with the parameters $\beta = 1.5$, $(\ql,\qr)=(1,0)$  for the same model as in Figure 1.  
The data 
agree with the results of our theory (\ref{mu}).
The inset shows ${\langle{\cal Q}^n\rangle_c / \tau_m}\bigr|_{ \tau_m \to \infty } $ 
which, as predicted by (\ref{mu}) , agree with the exact value of average current $J$. Numerical errors are smaller than the point-sizes.}
\label{fig2}
\end{figure}

 Since the particles are independent the current fluctuations can be described by a   Poissonian process
characterized by the rate at which 
walkers injected  (at rate $p_L$) from the left reservoir   end up  
(either after a  non stop flight or  a non direct flight) into the right reservoir or walkers injected (at rate $p_R$) from the right reservoir end up   into the left reservoir (see supplementary material \cite{suppl}). 
The current and its fluctuations can then be obtained by considering this process.
In the case $\ql=1$ and $\qr=0$ let ${\cal P}$ be the rate at which walkers,  which will end up into the right reservoir, are  injected from the left reservoir.  For general $\ql$ and $\qr$, because the walkers are independent and because they have no preferred direction, one has 
$ p_ L= \ql {\cal P}$ and $p_R= \qr {\cal P}$. 
We can then  write the characteristic function of  the integrated current  
$\mathcal{Q}=\int_{0}^{ {{\tau_m}} } dt J(t)$  
in measurement time $\tau_m$.  For $\tau_m$ much larger than the average residence time of a walker inside the system we get:
\bea
&& Z(\lambda) = \la e^{\lambda \mathcal{Q}} \ra 
\to \Pi_{t} \la e^{\lambda J(t) dt} \ra \nn \\
&&~~~~= \, \Pi_t [(1-p_L dt-p_R dt) + e^{\lambda} p_L dt + e^{-\lambda} p_R dt] \nn \\
&&~~~~=  e^{\mu (\lambda) {\tau_m} }~, \nn \\ 
&&{\rm where} ~~\mu (\lambda) = p_L (e^\lambda -1) { +} p_R (e^{-\lambda} { -} 1)~. \label{musol}
\eea
Hence we get  for the cumulant generating function of ${\cal Q}$ 
\bea
\mu(\lambda)= J~ \f{\ql (e^\lambda -1) +\qr (e^{-\lambda}-1)}{\ql-\qr} ~,\label{mu}
\eea
{where $J= (\ql-\qr) {\cal P}$.}
We can check that $\mu'(\lambda=0)$ gives the current $J$. Since in Eq.~(\ref{mu}) the system-size occurs 
only through the average current $J$,  
all cumulants of ${\cal Q}$ 
have  the same size dependence  as  the average current. This is consistent with recent numerical  simulations of  the 1D hard-point alternate mass gas \cite{brunet10}.
In Fig.(\ref{fig2}) we  check  the validity of Eq.~(\ref{mu}) from direct simulations of the open system. We discretized space and time and the reservoirs were taken large enough so that a steady state regime was achieved.

Not surprisingly  we also note that the following fluctuation theorem symmetry relation \cite{gc} is satisfied:
\bea
\mu (\lambda)= \mu[-\lambda -(\ln \ql -\ln \qr)]~.
\eea
\section{Current fluctuations in ring geometry}
In the ring geometry, the system consists  of  a fixed number $N$  of particles  which perform independent Levy walks  on a ring of length $L$. In the steady state, the density of particles is uniform.
As the walkers are independent, the  cumulants of the integrated current $ {\cal Q} $   are related to those of  the displacement  $x(t) $ of a single   walker on the infinite line (in the steady state)
\bea
\la{\cal Q}^n\ra_c  \sim 
{ N \over L^n} \la{ x(t)}^n\ra_c  = {\rho \over  L^{n-1}}   \la{ x(t)}^n\ra_c  
\label{Qnxn}
\eea
 where $\rho$ is the density on the ring.
If the walkers perform on the ring the same Levy walks  as on the infinite line, the cumulants of  $ x(t) $ and therefore those of ${\cal Q}$  grow,
as in (\ref{xsq},{\ref{xqd}}), 
faster than linearly with time 
(same exponent but a different prefactor as, on the ring, the walker is  in  its  steady state  rather than  starting  a flight at $t=0$).

On the other hand suppose one  introduces a cut-off time $\tau_L \sim L^\delta$  in the distribution $\phi(\tau)$ (for example by arguing that
 $\tau_L $ should be of the order of $t^*$, the relaxation time corresponding to the shortest wave number on the ring $k=2 \pi/L$, then using the 
result $t^* \sim k^{-\beta}$ \cite{suppl} one gets that $\delta  = \beta$;
 one could alternatively argue that,  as for the open  geometry, the length of the flights    cannot exceed the  system size and therefore   $\delta=1$). 
With such a  cut-off $\tau_L$, 
the cumulants of ${\cal Q}$ 
 would   grow linearly in time  (\ref{dif-cum},\ref{Qnxn}), with an amplitude which depends on the system size and on the cumulant considered
\bea
{\la{\cal Q}^2\ra_{ c}  \over t} \sim   L^{(2-\beta) \delta-1}
 \ \ \ ; \ \ \ {\la{\cal Q}^4\ra_{ c}  \over t} \sim   L^{(4-\beta) \delta-3}
\label{Qnxn-1}
\eea
 In one-dimensional mechanical models such as hard-point gas and anharmonic  chains, energy transport is mediated by phonons which are 
weakly scattered. One can then think of these as performing Levy walks and indeed this picture  is consistent with  simulation data on energy diffusion \cite{denisov03,metzler04,cipriani05,lepri11}.
Here we now see that the cut-off  time $\tau_L$ 
 also gives  a possible explanation for the behavior seen in simulations on the ring of hard-point alternate gas  in figure 3 of \cite{brunet10},
  where the cumulants grow linearly in time with different system size dependence
($\la{\cal Q}^2\ra_{ c} / t \sim L^{-0.5}$ and $\la{\cal Q}^4\ra_{ c} /t \sim L^{0.5}$). Then one gets from (\ref{Qnxn}) 
\bea
 \beta \sim 5/3 ~~{\rm and}~~ \delta \sim 3/2 \, \nonumber
\eea
which leads  through (\ref{Jsol}) to   a  value
 $\alpha =1/3$ for the anomalous Fourier's law of  the  hard-point alternate gas in the open geometry
consistent with most of the simulations done so far \cite{dhar08,grassberger,brunet10} for this system.

\section{Discussion}
In this work we have studied  the Levy diffusion model of anomalous heat 
transport. 
We have computed the average current (\ref{Jsol}), the energy profile  (\ref{Qpsol},\ref{pq}) and the large deviation function of the integrated current (\ref{musol}) , in the open geometry, i.e. when the system is connected at its two ends to reservoirs.  
One remarkable result is that all the cumulants of the integrated current have the same  anomalous size dependence  for the open geometry.
We have also proposed a simple   possible  explanation for  the  size dependence of the cumulants for the ring geometry.
 An interesting question would be to see how one could adapt existing theories \cite{LLP03,dhar08,beijeren,narayan,delfini,lukkarinen,lepri,LLP,P,LW,BBO,BDLLO} on anomalous conduction, which usually focus on the Green Kubo formula and on the average current, to predict the higher cumulants of the current both in the open and in the ring geometry. Of course a challenging issue would be to know whether the picture which emerges from the present work, (Levy walkers with a cut-off time in the ring geometry) could be confirmed by these theories.

 We thank NORDITA where this work was 
initiated. AD and KS thank S. Denisov for many useful discussions. AD thanks DST for support through the Swarnajayanti fellowship. KS was supported by MEXT (23740289).

\pagebreak
\renewcommand{\baselinestretch}{2}
\begin{widetext}

\section{{\bf Supplementary material for  ``Exact solution of a Levy walk model for  anomalous heat transport''}}
\vskip 1cm

The precise definition of the Levy walk model that we consider here is as follows. For a single particle each step of the walk consists in choosing a time of 
flight $t$ from the distribution $\phi(t)$ and then moving at  speed $v$ over a distance $x=vt$ in either direction, with equal probability. At any given 
time, the particle could either have landed at a point $x$ or could be passing over  that  point.    Accordingly let $Q(x,t) dx dt$ be the probability that a particle has precisely landed 
in the interval $(x,x+dx)$ during the time interval $(t,t+dt)$, and  let 
$P(x,t) dx$ denote the probability 
that the particle is in the interval  $(x,x+dx)$ at time $t$. Thus $P(x,t)$ 
includes events where the particle is crossing the 

interval $dx$. 
\\
We also define  
\begin{equation}
\psi(t)=\int_t^\infty d\tau~\phi(\tau) \label{psidef}
\end{equation}
 as the probability of choosing a time of flight $\geq t$ and
\begin{equation}
\chi(t)=\int_t^\infty d\tau~\psi(\tau)~.\label{chidef}
\end{equation}

{\bf Levy diffusion of a single particle on the infinite line}: 
For a particle starting from the origin $x=0$ at time $t=0$, the 
probability $P(x,t)$ satisfies 
\bea
 P(x,t)= {1 \over 2}\psi(t) \delta(|x|- v t )  \label{P(x,t)}
 + {1 \over 2} \int_0^t   d \tau \phi(\tau)  [P(x- v \tau,t-\tau) + P(x+ v \tau,t-\tau) ]~.
\nn
\eea
Taking the Fourier Laplace transform $\widetilde{P}(k,s)=\int_{-\infty}^\infty dx \int_0^\infty ~dt ~P(x,t) ~e^{i k x-st}$ we get
\bea
\widetilde{P}(k,s)=\f{\widetilde{\psi}(s-ikv)+\widetilde{\psi}(s+i k v)}{2-\widetilde{\phi}(s-i k v)-\widetilde{\phi}(s+i k v)}~,
\eea 
where $\widetilde{\phi}(s)=\int_0^\infty dt e^{-st} \phi(t)$ and 
$\widetilde{\psi}(s) =\int_0^\infty dt e^{-st} \psi(t) =[1-\widetilde{\phi}(s)]/s$. 
 Analysing the small $k$ and $s$ behavior of $\widetilde{P}(k,s)$
allows one to obtain formulae (4), (6) and (7) of the main paper.

{\bf Relaxation of density fluctuations:}
From   the evolution  equation for $P(x,t)$ [Eq.(3) in main text],  one can  see that  a density fluctuation of wave number $k$ relaxes  exponentially with a time constant $t^*$  obtained from the solution of   
$\int_0^\infty d \tau \phi(\tau) \cos (k {{v}} \tau) \exp[\tau/t^*] =1$.  For $\phi(\tau)$ of the form  given by Eq.~(5) in main text  one gets for small $k$ 
\bea
\label{t0}
1/t^* \simeq  A \Gamma(-\beta) \cos(\pi(1-\beta/2)) ~ {{(kv)^\beta}} {{/\langle \tau \rangle}}
\eea
whereas when  $\phi(\tau)$ has a finite  $\langle \tau^2 \rangle$, the regime is diffusive with $t^* \sim k^2$.

{\bf Levy diffusion in a finite system connected to infinite reservoirs:} 
In this case we consider our system to be the finite segment between $(0,L)$ and this is connected on the two sides to reservoirs.
The left reservoir consists of the region $  x \le 0$ while the right 
reservoir consists of the region $x \geq L $. We set $Q(x,t)=Q_l$ for 
points on the left reservoir and $Q(x,t)=Q_r$ for those on the right. 
In general if we know the distributions $Q(x,\tau)$ and $P(x,\tau)$ for all times 
$-\infty< \tau <t$ then the distribution at time $t$ is given by:
\bea
Q(x,t) =  \int_{-\infty}^\infty  dy \f{1}{2v} ~ Q(y,t-|x-y|/v) ~\phi(|x-y|/v)~, \label{Qeq} \\
P(x,t) = \int_{-\infty}^\infty  dy\f{1}{2 v}~ Q(y,t-|x-y|/v) ~\psi(|x-y|/v)~. \label{Peq}
\eea
In the above expressions $Q(x,t)$ gets contributions from walkers starting from
all possible points $y$ and landing precisely at $x$ at time $t$.  On the other hand $P(x,t)$ gets contributions  from walkers starting at $y$ and being either at or  passing $x$ at time $t$.
Since the distribution $Q(x,t)$ is constrained to take either of the values $Q_l$
or $Q_r$ in the reservoirs, the above equation gives, 
for points on the system
\begin{align}
Q(x,t) & = \f{Q_l}{2}~ \int_{-\infty}^0 dy \f{1}{2v}\phi [(x-y)/v]
+\f{Q_r}{2} ~\int_L^{\infty} dy \f{1}{2v}\phi [(y-x)/v] \nn \\
& + \int_{0}^L  dy \f{1}{2v} ~ Q(y,t-|x-y|/v) ~\phi(|x-y|/v)~,  \nn \\
& = \f{Q_l}{2}~ \psi(x/v) 
+\f{Q_r}{2} ~\psi[(L-x)/v] 
+ \int_{0}^L  dy \f{1}{2v} ~ Q(y,t-|x-y|/v) ~\phi(|x-y|/v)~, \nn \\
P(x,t) &= \f{Q_l}{2} ~\int_{-\infty}^0 dy \f{1}{2v}\psi [(x-y)/v]
+\f{Q_r}{2} ~ \int_L^{\infty} dy \f{1}{2v}\psi [(y-x)/v] \nn \\
&+ \int_{0}^L  dy\f{1}{2 v}~ Q(y,t-|x-y|/v) ~\psi(|x-y|/v)~, \nn \\
&= \f{Q_l}{2} ~ \chi(x/v) 
+\f{Q_r}{2} ~ \chi[(L-x)/v] + \int_{0}^L  dy\f{1}{2 v}~ Q(y,t-|x-y|/v) ~
\psi(|x-y|/v)~, \nn 
\end{align}
where we used the definitions of $\psi$ and $\chi$ from Eqs.~(\ref{psidef},\ref{chidef}).
In the steady state we have $Q(x,t)=Q(x)$ and $P(x,t)=P(x)$, hence we get:
\begin{align}
& Q(x)-\int_0^L  dy~\f{1}{2v} \phi(|x-y|/v) ~Q(y) =  \f{Q_l}{2} \psi(x/v)
+ \f{Q_r}{2} \psi[(L-x)/v]~,  \label{QeqSS} \\ 
& P(x) = \int_0^L  dy \f{1}{2v} \psi(|x-y|/v)~ Q(y)  
+  \f{Q_l}{2} \chi(x/v) + \f{Q_r}{2} \chi[(L-x)/v] ~. \label{PeqSS}
\end{align}
The solution of Eq.~(\ref{QeqSS}) is given by 
\begin{equation}
Q(x)=(Q_l-Q_r) H(x)+ Q_r \label{Qsol}
\end{equation}
where $H(x)$ is the probability that a Levy walker starting at position $x$ 
will first hit the left reservoir before it hits the right reservoir, and 
satisfies 
\begin{equation}
H(x)-\int_0^L  dy~\f{1}{2v} \phi(|x-y|/v) ~H(y) =  \f{1}{2} \psi(x/v)~.
 \label{Heq} 
\end{equation}

{\bf Steady state current. }
We re-write  Eq.~(\ref{Peq}) in the form 
\bea
P(x,t)=\int_{-\infty}^t d\tau [~Q(x-vt+v \tau,\tau) +Q(x+vt-v\tau,\tau)~]~\psi(t-\tau)/2~.\nn
\eea
 Taking a time-derivative and using the continuity equation $\p P(x,t)/\p t+\p J(x,t)/\p x=0$ we then obtain the following form of the current operator
\bea
 J(x,t) = \f{1}{2}\int_{-\infty}^{\infty} dy  ~ Q(x-y,t-|y|/v)  ~{\rm Sgn} (y) \psi(|y|/v) ~. \label{currexp}
\eea
This equation is easy to understand physically. 
The contribution to the integral coming from $y >0$ corresponds to 
particles crossing the point $x$ from left to right  which started their flight at  $x-y$ at time $t-y/v$ (the factor  $\psi(y/v)$  comes from the fact that these particles  have a flight time longer than $y/v$). Similarly 
the other part of the integral (from $y<0$) corresponds to a right-to-left 
current. 

In the steady state, setting $Q(x,t)\equiv Q(x)$  we get the result 
\bea
 J(x) = \f{1}{2}\int_{-\infty}^{\infty} dy  ~ Q(x-y)  ~{\rm Sgn} (y) \psi(|y|/v) ~. \label{sscurr}
\eea
Using the values of $Q$ in the reservoirs and the 
steady state solution given by Eqs.~(\ref{Qsol},\ref{Heq}) we evaluate the 
curent at $x=0$ and obtain
\bea
J=\f{(Q_l-Q_r)}{2}~\left[ ~\int_0^\infty dy \psi(y/v) - \int_0^L d y H(y) \psi(y/v)~\right]~. \label{jeq}
\eea

Since the system is non-interacting, this result can be obtained directly by 
noting   that the current is due to particles which  enter from the left
and leave to the right ( and to  the symmetric contribution). 
We then simply need to know the rate at which the non-interacting particles 
enter the system on the left side and leave the system into the right 
reservoir. This is given by
\bea
p_l=\f{Q_l}{2}\int_{-\infty}^0 dy~\int_{(L-y)/v}^\infty d \tau \phi(\tau) + 
\f{Q_l}{2} \int_{-\infty}^0 dy \int_{0}^{L} \f{dx}{v} ~[1-H(x)]~\phi[(x-y)/v]~, 
\eea
with a similar  expression for the right to left rate $p_r$. The net current given by $J=p_l-p_r$ is easily seen to be identical to Eq.~(\ref{jeq}).

{\bf Additivity principle:} An interesting observation is that 
the  generating 
function of the integrated current $\mu(\lambda)$ [given by Eq.~(19) in main text] matches exactly 
with the formula obtained from
 the  additivity principle (AP) \cite{BD04}  which 
 gives an expression for $\mu_{AP}(\lambda)$ in terms of the  the conductivity $D$ and equilibrium current 
fluctuations $\sigma$ defined respectively as 
\bea
\begin{array}{l}
D(Q)=\lim_{\Delta Q \to 0} L J  / \Delta Q ~,  \\
\\
\sigma(Q)= L ~{\lim_{t \to \infty} \la {\mathcal {Q}^2}\ra / t} ~.
\end{array}
\eea
The  expression for $\mu(\lambda)$ from AP is
\bea
\mu_{AP}(\lambda) &=& -\f{K}{L}\left[ \int_{\qr}^{\ql} dQ \f{D(Q)}{\sqrt{1+2 K \sigma(Q) }}\right]^2~, \nn \\
{\rm with}~~\lambda &=&\int_{\qr}^{\ql} dQ \f{D(Q)}{\sigma(Q)} 
\left[ \f{1}{\sqrt{1+2 K \sigma(Q) }} -1 \right]~.~~~~\label{muAP}
\eea
(this is a parametric expression: as $K$ varies, $\mu$ and $\lambda$ vary). 
From our exact results for $\mu(\lambda)$ we find $D=L p$ and $\sigma=L~\mu''(\lambda=0)=2 D Q~$. Using these in Eq.~(\ref{muAP}) and after explicitly 
performing the integrals we find $\mu_{AP}(\lambda)= \mu(\lambda)$. This result is somewhat surprising since the additivity principle is expected normally to hold 
for diffusive systems (here $D$ and $ \sigma$ have a $L$-dependence, whereas in usual diffusive systems they don't).

\end{widetext}

\end{document}